\documentclass[4pt,preprint]{emulateapj}

\def\kh{KH~15D}
\def\ltsima{$\; \buildrel < \over \sim \;$}
\def\lsim{\lower.5ex\hbox{\ltsima}}
\def\gtsima{$\; \buildrel > \over \sim \;$}
\def\gsim{\lower.5ex\hbox{\gtsima}}

\begin{document}

\title{The History of the Mysterious Eclipses of KH~15D\\
II. Asiago, Kiso, Kitt Peak, Mt.\ Wilson, \\
Palomar, Tautenburg and Rozhen Observatories, 1954-97 }
         
\author{
John Asher Johnson\altaffilmark{1},
Joshua N.\ Winn\altaffilmark{2,3},
Francesca Rampazzi\altaffilmark{4},
Cesare Barbieri\altaffilmark{4},
Hiroyuki Mito\altaffilmark{5},
Ken-ichi Tarusawa\altaffilmark{5},
Milcho Tsvetkov\altaffilmark{6,7},
Ana Borisova\altaffilmark{7},
Helmut Meusinger\altaffilmark{8}
}

\altaffiltext{1}{Department of Astronomy, University of California,
Mail Code 3411, Berkeley, CA 94720}

\altaffiltext{2}{Harvard-Smithsonian Center for Astrophysics, 60
Garden St., Cambridge, MA 02138}

\altaffiltext{3}{Hubble Fellow}

\altaffiltext{4}{Department of Astronomy, University of Padova, Vicolo
Osservatorio 2, 35122 Padova Italy}

\altaffiltext{5}{Kiso Observatory, Institute of Astronomy, School of
Science, University of Tokyo, Mitake-mura, Kiso-gun, Nagano-ken,
397-0101, Japan}

\altaffiltext{6}{Institute of Astronomy, Bulgarian Academy of
Sciences; Tsarigradsko Shosse Blvd. 72, Sofia-1784, Bulgaria}

\altaffiltext{7}{Space Research Institute, Bulgarian Academy of
Sciences, Moskovska Str. 6, Sofia-1000, Bulgaria}

\altaffiltext{8}{Th\"{u}ringer Landessternwarte Tautenburg, Sternwarte
5, D-07778 Tautenburg, Germany}

\begin{abstract}
The unusual pre--main-sequence binary star named \kh\ undergoes
remarkably deep and long-lasting periodic eclipses. Some clues about
the reason for these eclipses have come from the observed evolution of
the system's light curve over the last century. Here we present $UBVRI$
photometry of KH~15D based on photographic plates from various
observatories, ranging in time from 1954 to 1997. The system has been
variable at the $\approx$1~mag level since at least 1965. There is no
evidence for color variations, with a typical limit of
$\Delta(B-V)<0.2$~mag. We confirm some previously published results
that were based on a smaller sample of plates: from approximately 1965
to 1990, the total flux was modulated with the 48-day orbital period
of the binary, but the maximum flux was larger, the fractional
variations were smaller, and the phase of minimum flux was shifted by
almost a half-cycle relative to the modern light curve. All these
results are consistent with the recently proposed theory that \kh\ is
being occulted by an inclined, precessing, circumbinary ring.
\end{abstract}


\section{Introduction}

The T~Tauri star known as \kh\ \citep{kearns98} called attention to
itself through a remarkable pattern of photometric variations.  Every
48 days, the system dims by over 3 magnitudes \citep{hamilton01,
herbst02}.  It remains in this faint state for a duration that has
grown from about 16 days in 1997, when the periodicity of the
variations was first noticed, to over 24 days in 2004.  Before 1999,
the faint state was characterized by a short interruption of about 5
days during which the system returned to its bright state.  Since
then, these ``rebrightening'' events have subsided to
near-undetectability.

\citet{hamilton01} and \citet{herbst02} argued persuasively that these
dramatic and unusual variations are due to eclipses by circumstellar
dust that surrounds the visible K7 star, although the arrangement of
the dust and the cause of the periodicity were unknown.  At the time,
it was not even 
clear whether the system was a single star or a multiple star system,
and hence whether the relative velocity of the dusty structure and the
visible star was due to the orbital motion of the dust, or of the
star \citep{herbst02}.  The mystery of the eclipses of \kh\ motivated
an intensive monitoring campaign by observers around the world
\citep{herbst02, barsunova04}, attracted other observers seeking clues
in deep imaging \citep{tokunaga04} and spectroscopy \citep{agol04,
deming04}, and also attracted theorists hoping to model the
circumstellar environment \citep{barge03, grinin02}.

An important step in the clarification of the basic system properties
was the analysis of archival photographic plates of NGC~2264, the
3~Myr old star cluster in which KH~15D resides. Photographs from years
prior to 1951 showed no evidence of eclipses deeper than 1~mag
\citep{winn03}. More interestingly, photometry from 1967--1982
\citep[][hereafter Paper I]{johnson04} showed that eclipses were
occurring with the same 48-day period as today, but with three
striking differences: the system was brighter at all phases, the
fractional variations in the total light were smaller, and the phase
of minimum light was shifted relative to the modern light curve.

Motivated by these findings, \citet{winn04} proposed that \kh\ is a
binary star that is gradually being occulted by an opaque foreground
screen, which is probably the edge of an inclined, precessing,
circumbinary disk. \citet{chiang04} had the same idea independently,
and explored the dynamical issues further, arguing that the ``disk''
was likely to be a fairly narrow ring. A radial velocity study by
\citet{johnson04b} confirmed that \kh\ is a binary star, and to the
extent that the orbital parameters of the binary could be constrained
by the radial velocities, the orbital parameters agreed well with the
predictions of \citet{winn04}. These developments have made \kh\ less
mysterious than it once was, and they bring closer to reality the
prospect of using the system to learn about disk evolution, the
circumstellar environment of young stars, and possibly even planet
formation.

A prerequisite for a complete model of the binary orbit and the
surrounding disk or ring is a light curve with reasonably complete
time and phase coverage over the entire era of the periodic
variations. Towards that end, we present additional photometry of \kh\
based on an analysis of photographic plates from several archives
around the world, ranging in time from 1954 to 1997, with most of the
data coming from years prior to 1985.  Our plate selection and
digitization process are described in \S~\ref{plates}, and the
photometric measurements and uncertainties are described in
\S~\ref{photometry}.  In \S~\ref{results}, we present the light curves
and describe the evolution that have been observed over the past 50
years. In \S~\ref{discussion} we discuss these findings in the context
of our model of the system.

\section{The plates}
\label{plates}

In Paper I, we presented results from an analysis of 52 photographic
plates in the archive of the Astrophysical Observatory of Asiago, in
northern Italy. All of those plates were derived from observations
with a single telescope, the 67/92~cm Schmidt telescope, and almost
all of them were exposed with a red filter (RG5, or equivalently
RG665) and a red-sensitive emulsion (I--N). We deliberately selected
this subset of plates for the first phase of our archival study of
\kh, because they formed a fairly dense time series with uniform
characteristics, and because the effective band pass was approximately
the same as the Cousins $I$ band, in which most of the modern
CCD-based photometry has been obtained.  In addition, the photometry
was simplified for the I--N/RG5 plates because the glare from a nearby
B star is much smaller on red-sensitive plates than it is on the more
commonly used blue-sensitive plates.

Now that our photometric procedure has been tested thoroughly, we have
expanded our analysis to include a heterogeneous collection of plates,
obtained with different telescopes and effective band passes. The
sample of 87 plates that is considered in this paper is drawn from the
following sources\footnote{Most of these plates were identified by
querying the Wide Field Plate Database, http://www.skyarchive.org}:

\begin{enumerate}

\item There are 26 plates from Asiago Observatory, the same archive
from which the sample in Paper I was drawn \citep{barb03}. But whereas
the previously analyzed plates were taken with the 67/92~cm Schmidt
telescope, the plates in the present sample were taken with the
40/50~cm Schmidt telescope and the 1.22~m Galilei telescope.  As in
Paper I, these plates were digitized at Asiago Observatory with a
1600~dpi, 14-bit commercial flatbed scanner.

\item There are 35 plates that were taken with the 105/150~cm Schmidt
telescope at Kiso Observatory, in Japan, a facility of the University
of Tokyo.  The Kiso plates were digitized at the observatory with a
PDS microdensitometer.

\item There are 7 plates that were taken with the 134/200~cm Schmidt
telescope of the Th\"uringer Landessternwarte Tautenburg, which is
located near Jena, in Germany. They were digitized using the local
high-precision plate scanner TPS at 2540 dpi and 12 bits~pixel$^{-1}$.
\citep{brunzendorf99}.

\item There are 7 plates that were taken with the 4~m Mayall telescope
at Kitt Peak National Observatory, which is located near Tucson,
Arizona. The plates were shipped to the Harvard College Observatory
for digitization with a 1600~dpi, 14 bits~pixel$^{-1}$ commercial
flatbed scanner.

\item There are 5 plates taken with the 60~inch telescope of Mount
Wilson Observatory, in the San Gabriel mountains of California, which
were also shipped to the Harvard College Observatory for scanning.

\item There are 4 plates that were taken with the 2~m
Ritchey-Chretien-Coude (RCC) telescope at the National Astronomical
Observatory in Rozhen, Bulgaria. These plates were scanned with a
1600~dpi, 14 bits~pixel$^{-1}$ scanner that is similar to the one used
at Asiago Observatory.

\item Finally, there are 3 plates from the Palomar Observatory Sky
Survey obtained with the 122/183~cm Oschin Schmidt telescope. These
were scanned using the USNO PMM plate scanner. The digitized images
were downloaded from the Space Telescope Science Institute web
site\footnote{http://archive.stsci.edu/dss/}. 

\end{enumerate}

\begin{figure*}[!t]
\epsscale{0.75}
\plotone{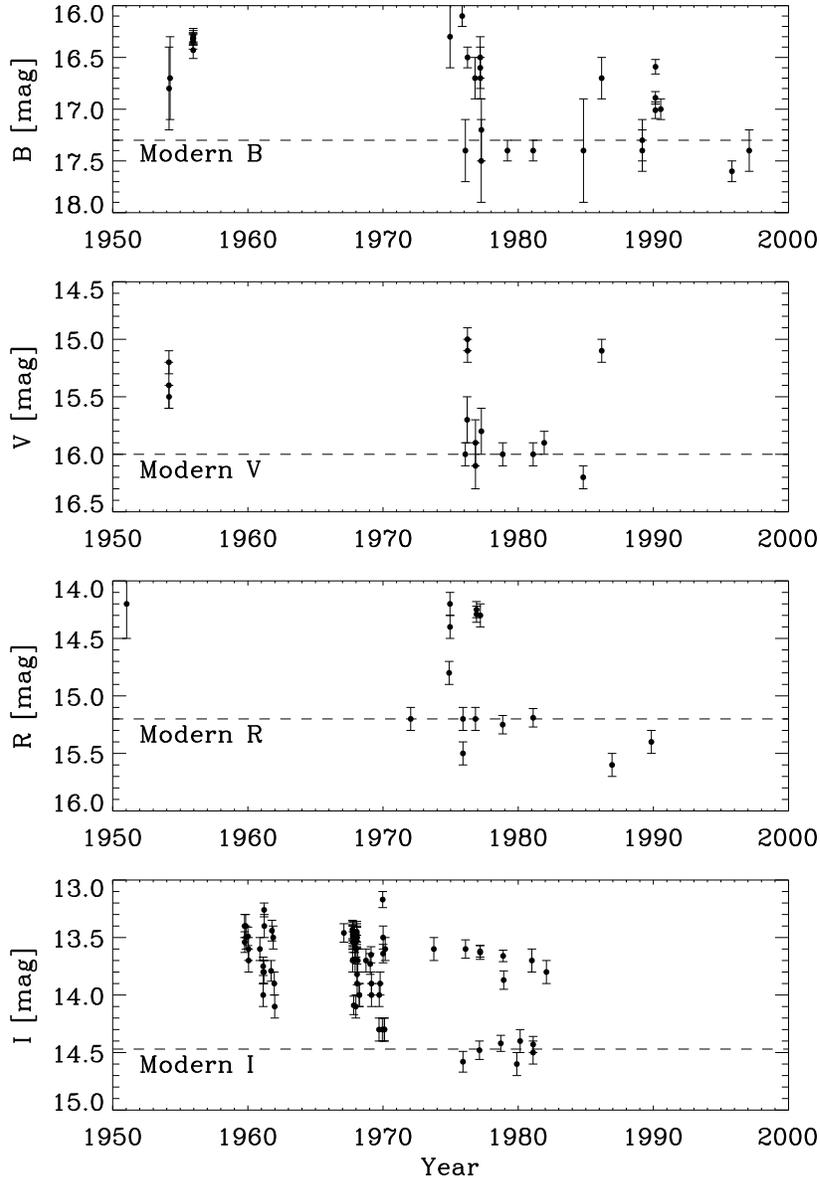}
\caption{ $BVRI$ photometry of \kh\ over the last 5
decades derived from archival photograhpic plates. \label{lightcurves} }
\end{figure*}

\section{Photometry}
\label{photometry}

The scanning process produces a digital record of the density of
opaque grains in the photographic emulsion as a function of the
position on the plate.  As in Paper~I, we used a form of profile
photometry (also known as point-spread-function, or PSF, photometry)
in order to estimate the apparent magnitude of \kh\ on each plate,
relative to a grid of local reference stars.  This procedure is
described fully in Paper~I and is summarized below.

The relationship between the density enhancement, $d(x,y)$, and the
flux received from the star, $S$, is nonlinear, and for bright stars
the density saturates at a maximum value. Thus, an accurate PSF model
must include a nonlinear response.  We employed the model described by
\citet{stetson79}, in which the PSF model is a magnitude-dependent
transformation of a two-dimensional Gaussian distribution,
\begin{equation}
d_S(x,y) = \left[ d_G(x,y)^{-q} + d_{\rm sat}^{-q} \right]^{-1/q},
\label{eq:pdp}
\end{equation}
where $d_{\rm sat}$ and $q$ describe the saturation properties of the
plate, and $d_G(x,y)$ is a Gaussian function of the coordinates $x$
and $y$.  The Stetson function has the desirable limits $d_S
\rightarrow d_G$ for $d_G \ll d_{\rm sat}$, and $d_S \rightarrow
d_{\rm sat}$ for $d_G \gg d_{\rm sat}$.

The model is fitted to the stellar image using a non-linear
least-squares algorithm, and $S$ is computed as the two-dimensional
integral of $d_G(x,y)$.  This flux scale is related to a standard
magnitude via the expression
\begin{equation}
m = m_0 - 2.5(1 + \epsilon) \log_{10}{S},
\label{mag_equation}
\end{equation}
where $m$ is the apparent magnitude in the effective band pass
produced by the emulsion and filter combination, and $\epsilon$
accounts for nonlinearity.

For each plate, we selected the band pass in the
Johnson-Morgan-Cousins $UBVRI$ system that is the closest match to the
emulsion/filter combination that was employed (see
Table~\ref{sumtable}). Then we estimated $m_0$ and $\epsilon$ for that
plate using the set of 53 reference stars that we identified in
Paper~I. Those stars were selected by virtue of their proximity (they
are all within a 20\arcmin~box centered on \kh), their wide range of
magnitudes and colors, and their relatively low level of variability.
After measuring $S$ for each of the reference stars, we found the
values of $m_0$ and $\epsilon$ that minimize the sum of the squared
residuals between Eqn.~\ref{mag_equation} and the magnitudes (in the
appropriate band pass) measured by \citet{flaccomio99}. In Paper~I, we
noted that higher order nonlinear terms and color corrections were
unhelpful in reducing the scatter in the magnitude relation.  We found
the same to be true for the present sample.

\begin{figure*}[!t]
\epsscale{1}
\plotone{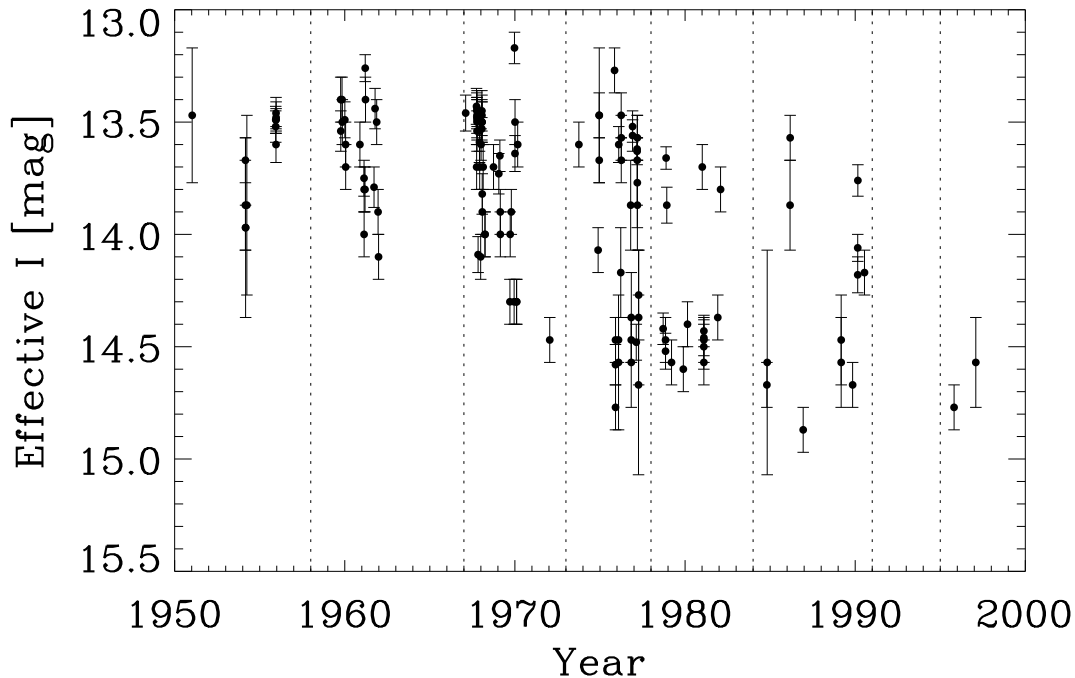}
\caption{ Effective $I$ band photometric time series of KH~15D, based on the
assumption that the system has always had the same color indices as
the modern uneclipsed state.\label{iband} }
\end{figure*}

For almost all the plates, these steps were carried out by an
automatic procedure, but there were a few plates that required
individual attention. Two of the Tautenburg plates had an effective
band pass approximating $U$ band, but \citet{flaccomio99} did not
measure $U$ band magnitudes of our reference stars; for those plates,
we used the $U-B$ measurements of \citet{park00}. On the Mt.\ Wilson
plates, there was a strong gradient in the background near \kh, due to
the glare of the nearby B star; for those plates, we found it
necessary to account for this gradient by fitting a first-order
two-dimensional polynomial surface to the background. Finally, visual
inspection of the POSSII~IR plate revealed that \kh\ was much fainter
than than any of our reference stars. Rather than extrapolating to
find the magnitude of \kh\ on this plate, we instead assign a firm
lower limit of $I > 15.5$ based on the magnitude of the faintest of
our reference stars.

\begin{figure*}[!t]
\epsscale{1.00}
\plotone{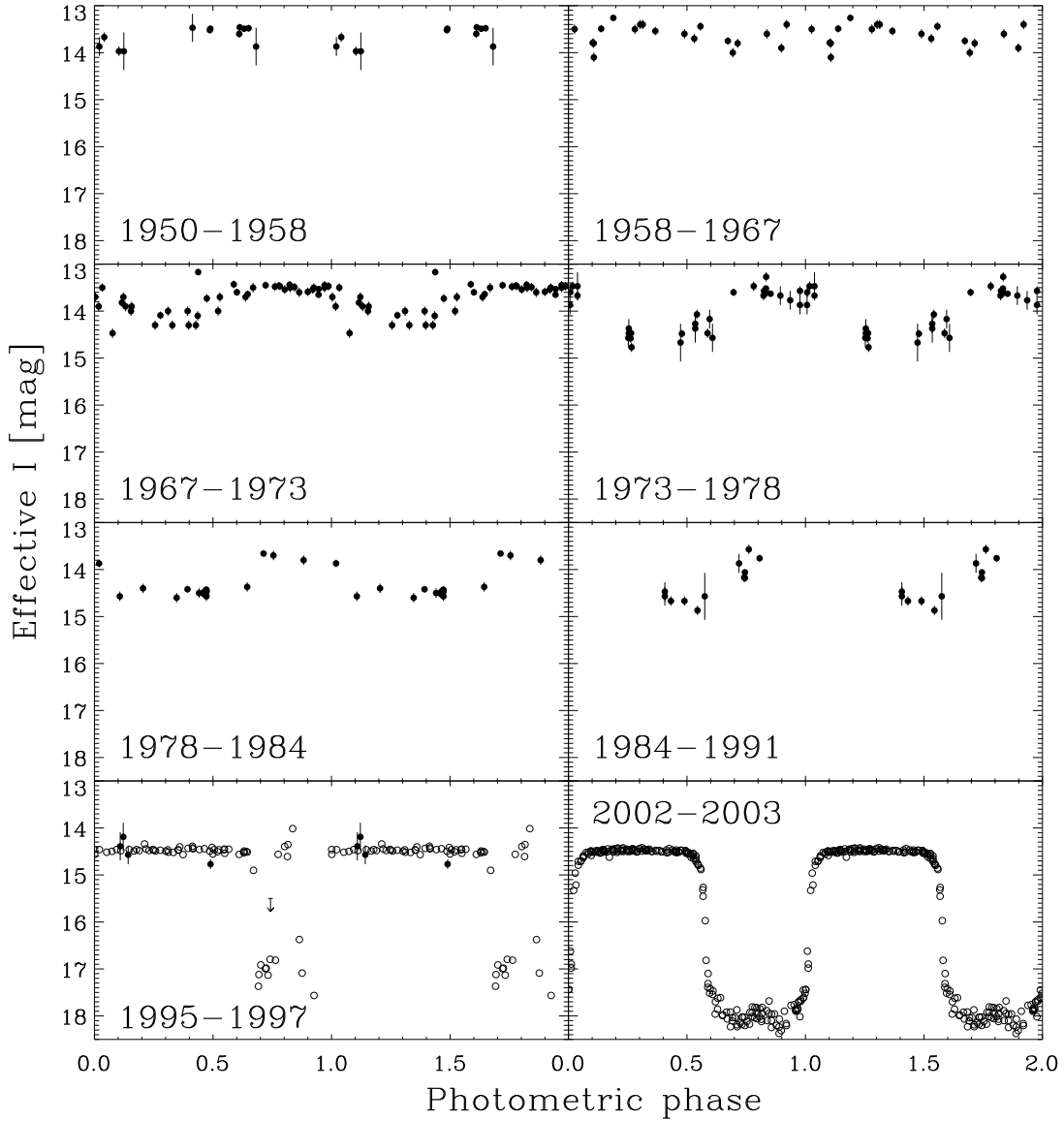}
\caption{ Phased light curves for different time periods. The filled
circles are data based on the photographic plates described in this
paper.  The open circles (in the bottom panels) are modern CCD data,
for which the error bars are smaller than the plot
symbols. \label{phased} }
\end{figure*}

\subsection{Estimation of Errors}

To estimate the uncertainty of each photometric measurement, we
followed the same procedure as in Paper~I.  For each subset of plates
obtained at the same telescope and with similar emulsion/filter
combinations, we plotted the standard deviation in the 
resulting magnitudes of each star {\it versus} the mean magnitude of
that star.  The lower envelope of that relationship was taken to be
the limiting uncertainty of our fitting procedure as a function of
magnitude, $\sigma_{I,min}(I)$.  We also took into account the
relative quality of each plate by computing $\sigma_m$, the scatter in
the fit to Eqn.~\ref{mag_equation}, and then we estimated the total
uncertainty as
\begin{equation}
{\rm Unc.} = \sigma_{I, {\rm min}}(I) \times 
             \left( \frac{\sigma_m} {\sigma_{m,{\rm min}}} \right),
\label{errors}
\end{equation}
\noindent where $\sigma_{m,{\rm min}}$ is the minimum $\sigma_m$ among
all plates in that subset.

\section{Results}
\label{results}

\subsection{$BVRI$ light curves}

The magnitudes of \kh\ are given in Table~\ref{datatable}. The first
four columns give the Julian Date, the observatory (using the
designations defined in Table~\ref{sumtable}), the closest
corresponding $UBVRI$ band pass, and the apparent magnitude with the
the corresponding estimated uncertainty.  The light curves for each
band pass are shown in Figure~1.  For completeness, we have also
plotted the measurements from Paper~I.

The $BVRI$ light curves confirm two of the findings of Paper~I: in the
past, the system was formerly brighter and the fractional variations
in total flux were smaller.  The horizontal dashed lines in Figure~1
show the modern uneclipsed magnitudes of \kh.  The light curves show
that before about 1990, the system was often as much as 1~mag brighter
than it appears today.

\subsection{Test for chromatic variations}

\citet{hamilton01} and \citet{herbst02} found that the flux variations
that are observed today are nearly achromatic.  Outside of eclipses,
the $V-I$ and $V-R$ color indices do not vary by more than about
0.05~mag.  Even during the 3~mag fading events, the system hardly
changes color, although at minimum light it does appear to be
systematically bluer by $\approx$0.1~mag in $V-R$.  It would be
interesting to know whether the flux variations in previous decades
were also nearly achromatic, or whether they were accompanied by color
changes.

The measurements reported in Paper~I were almost all in $I$ band.  We
considered only a few plates with different band passes, and only one
pair of plates with different band passes ($B$ and $R$) that were
taken nearly simultaneously (on 1974~December~15).  The result was
$\Delta(B-R)= -0.14\pm 0.30$, where $\Delta(B-R)$ is the difference
between the $B-R$ index measured on that date, and the $B-R$ index of
the modern uneclipsed state.

With the multi-band sample presented in this paper, we were able to
test for color changes more extensively. There are 18 pairs of
observations that took place within 2~days of each other and that have
different effective band passes. For each of those pairs,
Table~\ref{kh_colors} lists the mean Julian date and the measured
color index, $C$. The third column lists $\Delta C$, which the
difference between $C$ and the corresponding color index of the modern
uneclipsed state\footnote{Here, for consistency, we have used the
\citet{flaccomio99} values for the uneclipsed color indices. However,
following \citet{hamilton01}, we have assumed that the $B-V$ entry for
\kh\ (star no.\ 391) in the catalog of \citet{flaccomio99} contains a
typo, and should be 1.32 rather than 0.32. This makes the entry
consistent with other published measurements (e.g.\ Park et al.\ 2000,
who report $B-V=1.308$) and with the color expected of a star of the
observed K7 spectral class.}. The final column gives the photometric
phase, which is defined in \S~\ref{evo}.

The result of this analysis is that there were no detectable color
changes: all of the values of $\Delta C$ are consistent with zero,
regardless of the photometric phase.  Thus it appears as though the
optical color indices of \kh\ have remained constant to within
$\approx$0.2~mag over the past few decades.

\subsection{An ``effective'' $I$ band time series}

Under the assumption that the color indices have always been exactly
constant, we can put all the measurements on the same magnitude scale.
The last column of Table~\ref{datatable} gives the ``effective''
$I$-band magnitude, which was computed under the assumptions
$U-I=4.02$, $B-I=2.94$, $V-I=1.62$, and $R-I=0.80$. This effective $I$
band time series is shown in Figure~\ref{iband}.

\subsection{Evolution of the light curve}
\label{evo}
We have divided the effective $I$ band time series into different time
intervals, bounded by the dotted lines in Figure~\ref{iband}, in order
to examine how the time series has evolved over the past 5 decades.
Since both the modern variations and the variations observed from
1967-1982 (Paper~I) are nearly periodic, it is useful to plot the
magnitude as a function of the photometric phase $\phi$, which is
defined as
\begin{equation}
\phi = \frac{ ({\rm J.D.} - 2,434,800)~\mbox{mod}~P } {P},
\label{eph}
\end{equation}
where $P=48.38$~days is the orbital period that was determined by
\citet{johnson04b}.  We note that this definition of $\phi$ differs
from the photometric phase that was used in Paper~I, because of the
updated period.  In Figure~3 we have plotted phased light curves for
each of the 6 time intervals covered by our data, along with phased
light curves from CCD-based data that were kindly provided by W.\
Herbst and C.\ Hamilton.

The phased light curves clarify the nature of the photometric
variations.  From about 1965 to 1991, the system alternated between a
bright state and a faint state with a 48-day period, just as the
system does today.  However, the modern bright state is approximately
1~mag fainter than the pre-1991 bright state.  In fact, the modern
bright state has approximately the same brightness as the pre-1991
{\it faint} state.  The other striking feature of the phased light
curves is that $\phi_{\rm min}$, the phase corresponding to the
midpoint of the faint state, has changed by nearly 0.5.  Formerly it
occurred at $\phi_{\rm min} \approx 0.32$.  By 1995-96, it had shifted
to $\phi_{\rm min} = 0.8$, although it is worth remembering that in
that year and the few years to follow, the ``faint state'' included a
short-lived rebrightening event.

These findings confirm and extend the results of Paper~I, which were
based almost entirely on the 1965-73 data.  The enlarged sample
indicates that the shallow, phase-shifted eclipses were occurring
until at least 1991. It also shows that the variations were smaller
and less regular in the years before 1965. Between 1958 and 1965,
there is some evidence for a faint state centered at $\phi_{\rm min}
\approx 0.7$, but the fractional variation appears to be
$\lsim$0.5~mag, and the transition between the bright and faint states
is more gradual. Before 1958, no particular pattern of variations is
evident.

\section{Discussion}
\label{discussion}

\citet{winn04} and \citet{chiang04} hypothesized that \kh\ is a pair
of stars in a mutual orbit with high eccentricity, and the
``eclipses'' are actually occultations by a circumbinary disk. A full
analysis of this model should include all the available information,
including the modern photometry, the photographic photometry, and the
radial velocity measurements by \citet{johnson04b}. This will be the
subject of a future paper. Here we simply describe how the basic
properties of the light curves presented in this paper can be
explained by the model, and point out one aspect of the new data that
has not yet been explained.

In the model, we are viewing an eccentric binary system nearly
edge-on. Today we can see only one of the two stars; we refer to the
visible star as A, and the invisible one as B. The reason B is
invisible is that its entire orbit is in the shadow of a circumbinary
disk. Also shadowed is a portion of A's orbit, near periapse. The
3~mag eclipses occur when A passes into the shadow, once per
period. The reason for the decadal evolution of the light curve is
that the line of nodes of the disk is regressing, in response to the
gravitational torque produced by the stars. This causes the shadow of
the disk to sweep across the system.
As a progressively larger portion of A's orbit is shadowed,
the eclipses increase in duration. Extrapolating backwards in time,
star A used to be completely exposed, and only a portion of star B's
orbit (near apoapse) was shadowed. In that era, we observed periodic
occultations of star B.

Thus, we interpret the historical light curve as follows. The periodic
alternation between the bright and faint state is due to the periodic
occultation of star B. The total flux was formerly larger, and the
fractional variations were formerly smaller, because the steady light
of star A was visible at all orbital phases. Finally, the reason for
the phase shift is that the previous eclipses occurred around apoapse,
whereas the modern eclipses occur around periapse. 

Extrapolating even further backwards in time, one would expect to
reach an era in which neither star was shadowed at any phase, and
there were no photometric variations due to occultations. The new data
presented in this paper show a somewhat more complicated picture.
From 1951 to 1965, the system varied at the $\approx$0.5~mag level.
This is certainly smaller than the 1~mag periodic variations observed
between 1965 and 1991, but it is not zero. It is not clear whether
these early variations were periodic, or erratic. The 1958-1965 data do
seem to obey the 48-day periodicity, but the data are too sparse and
the variations are too small to be certain.

In closing, we wish to note that although we have attempted to find as
many useful photographic plates of NGC~2264 as possible, we do not
claim that our collection is complete. Undoubtedly there are
additional plates in other observatory archives. Additional photometry
would be useful for a quantitative determination of the orbital
parameters of the binary star and its putative circumbinary disk,
especially if those data were taken during the years 1980-1995 where
the time coverage of our light curves is not very complete.

\acknowledgments We are indebted to Paolo Maffei, Gino Totsi,
Yoshikazu Nakada, George Carlson, 
Steve Strom, Bill Schoening, Jean Mueller, Tony Misch, Peter Quinn,
and Dave Monet for their kind assistance with the plate archives. We
have enjoyed helpful conversations with Catrina Hamilton, Bill Herbst,
Matt Holman, Geoff Marcy, and Fred Vrba about this work. M. Bagaglia 
is to be thanked for the digitization of part of the plates
obtained by P. Maffei. We also thank Erin Johnson for assistance with
the chromatic analysis and Liliana Lopez for help with the visual
inspection of the Mt. Wilson plates. This work 
makes use of data from the digitized Italian photographic
archives, produced under contract MIUR/COFIN 2002 to C.\ Barbieri,
Department of Astronomy, University of Padova. J.N.W.\ is
supported by NASA through Hubble Fellowship grant HST-HF-01180.02-A,
awarded by the Space Telescope Science Institute, which is operated by
the Association of Universities for Research in Astronomy, Inc., for
NASA, under contract NAS 5-26555. M.T.\ acknowledges support from COST
Action-283 and grant BNSF I-1103/2001 of the Bulgarian Ministry of
Education and Science.

\clearpage

\begin{deluxetable}{lllcccccc}
\tabletypesize{\scriptsize}
\tablecaption{The plate sample \label{sumtable}}
\tablewidth{0pt}
\tablehead{
\colhead{Observatory} & 
\colhead{Name of} & 
\colhead{Range of} & 
\colhead{Number of} &
\colhead{} &
\colhead{} &
\colhead{Approximate} & 
\colhead{Scanning} \\
\colhead{(Designation)} & 
\colhead{Telescope} & 
\colhead{Years} & 
\colhead{Plates} &
\colhead{Filter} &
\colhead{Emulsion} &
\colhead{Band} & 
\colhead{Method} 
}
\startdata
Asiago (ASI50)   & 40/50~cm Schmidt & 1959-1969       & 20 & RG5, RG665 & I-N    & I & Flatbed \\
Asiago (ASI120)  & 1.22~m Galilei   & 1955            & 6  & none       & 103a-O & B & Flatbed \\
Kiso (KIS0)      & 105/150~cm Kiso Schmidt   & 1975-1986       & 11 & GG385      & IIa-O & B & PDS     \\
\nodata          & \nodata          & \nodata         & 7  & GG495      & 103a-D & V & \nodata \\
\nodata          & \nodata          & \nodata         & 4  & GG495      & IIa-D  & V & \nodata \\
\nodata          & \nodata          & \nodata         & 1  & RG610      & 103a-F & R & \nodata \\
\nodata          & \nodata          & \nodata         & 2  & RG610      & IIa-F  & R & \nodata \\
\nodata          & \nodata          & \nodata         & 4  & RG645      & 103a-E & R & \nodata \\
\nodata          & \nodata          & \nodata         & 6  & RG695      & I-N    & I & \nodata \\
Kitt Peak (KPNO) & 4.0~m Mayall     & 1978-1981       & 2  & RG610      & IIIa-F & R & Flatbed \\
\nodata          & \nodata          & \nodata         & 2  & RG695      & IV-N   & I & \nodata \\
\nodata          & \nodata          & \nodata         & 1  & GG385      & IIIa-J & V & \nodata \\
\nodata          & \nodata          & \nodata         & 1  & GG495      & IIa-O  & B & \nodata \\
\nodata          & \nodata          & \nodata         & 1  & GG385      & IIIa-J & B & \nodata \\
Mt.\ Wilson (MTW)& 60~in            & 1954            & 3  & GG11       & 103a-D & V & Flatbed \\
\nodata          & \nodata          & \nodata         & 2  & GG13       & 103a-O & B & \nodata \\
Palomar (POSS) & 122/183~cm Oschin Schmidt   & 1951-1996   & 1  & RP2444     & 103a-E  & R & PMM \\
\nodata & \nodata & \nodata         & 1  & RG610     & IIIa-F  & R & \nodata \\
\nodata & \nodata & \nodata         & 1  & RG9     & IV-N  & I & \nodata \\
Rozhen NAO (ROZ) & 2.0~m RCC         & 1990            & 4  & GG385      & ZU21   & B & Flatbed \\   
Tautenburg (TAU) & 134/200~cm Schmidt & 1972-1997       & 4  & GG13       & ZU     & V & TPS     \\
\nodata          & \nodata          & \nodata         & 2  & UG2& ZU & U & \nodata \\
\nodata          & \nodata          & \nodata         & 1  & RG1        & 103a-E & R & \nodata \\
\enddata
\end{deluxetable}

\begin{deluxetable}{ccccc}
\tablecaption{Photometric Measurements of KH~15D \label{datatable}}
\tablewidth{0pt}
\tablehead{
\colhead{Julian Date} & 
\colhead{Observatory} & 
\colhead{Band Pass} & 
\colhead{$m$ [mag]} & 
\colhead{$I_{\rm eff}$ [mag]\tablenotemark{a}} 
}
\startdata
2433658.9 & POSS & R & $14.2 \pm 0.4$ & 13.4 \\
2434801.0 & MTW & V & $15.4 \pm 0.2$ & 13.7 \\
2434802.0 & MTW & V & $15.2 \pm 0.2$ & 13.5 \\
2434805.0 & MTW & V & $15.5 \pm 0.1$ & 13.9 \\
2434806.0 & MTW & B & $16.8 \pm 0.4$ & 13.9 \\
2434833.0 & MTW & B & $16.7 \pm 0.4$ & 13.7 \\
2435452.5 & ASI120 & B & $16.35 \pm 0.08$ & 13.41 \\
2435452.6 & ASI120 & B & $16.32 \pm 0.07$ & 13.38 \\
2435458.5 & ASI120 & B & $16.43 \pm 0.08$ & 13.49 \\
2435458.6 & ASI120 & B & $16.29 \pm 0.08$ & 13.35 \\
2435459.5 & ASI120 & B & $16.32 \pm 0.06$ & 13.38 \\
2435460.4 & ASI120 & B & $16.31 \pm 0.07$ & 13.37 \\
2436846.6 & ASI040 & I & $13.4 \pm 0.2$ & 13.4 \\
2436849.7 & ASI040 & I & $13.5 \pm 0.10$ & 13.5 \\
2436876.5 & ASI040 & I & $13.4 \pm 0.1$ & 13.4 \\
2436881.6 & ASI040 & I & $13.5 \pm 0.1$ & 13.5 \\
2436935.4 & ASI040 & I & $13.49 \pm 0.09$ & 13.49 \\
2436952.4 & ASI040 & I & $13.6 \pm 0.1$ & 13.6 \\
2436954.4 & ASI040 & I & $13.7 \pm 0.1$ & 13.7 \\
2437259.5 & ASI040 & I & $13.6 \pm 0.2$ & 13.6 \\
2437348.3 & ASI040 & I & $13.75 \pm 0.08$ & 13.75 \\
2437349.3 & ASI040 & I & $14.0 \pm 0.1$ & 14.0 \\
2437350.3 & ASI040 & I & $13.8 \pm 0.1$ & 13.8 \\
2437369.3 & ASI040 & I & $13.8 \pm 0.1$ & 13.8 \\
2437373.3 & ASI040 & I & $13.26 \pm 0.07$ & 13.26 \\
2437379.3 & ASI040 & I & $13.4 \pm 0.1$ & 13.4 \\
2437562.6 & ASI040 & I & $13.8 \pm 0.10$ & 13.8 \\
2437584.6 & ASI040 & I & $13.44 \pm 0.09$ & 13.44 \\
2437619.6 & ASI040 & I & $13.5 \pm 0.1$ & 13.5 \\
2437649.5 & ASI040 & I & $13.9 \pm 0.1$ & 13.9 \\
2437659.6 & ASI040 & I & $14.1 \pm 0.2$ & 14.1 \\
2440578.4 & ASI040 & I & $13.17 \pm 0.07$ & 13.17 \\
2441335.0 & TAU & R & $15.2 \pm 0.1$ & 14.4 \\
2442726.3 & KISO & B & $16.1 \pm 0.2$ & 13.2 \\
2442747.0 & KISO & I & $14.6 \pm 0.10$ & 14.6 \\
2442747.1 & KISO & R & $15.2 \pm 0.1$ & 14.4 \\
2442747.2 & KISO & R & $15.5 \pm 0.1$ & 14.7 \\
2442811.1 & KISO & V & $16.0 \pm 0.2$ & 14.3 \\
2442812.1 & KISO & B & $17.4 \pm 0.3$ & 14.5 \\
2442859.9 & KISO & V & $15.7 \pm 0.2$ & 14.1 \\
2442868.9 & KISO & V & $15.0 \pm 0.1$ & 13.4 \\
2442870.9 & KISO & B & $16.5 \pm 0.2$ & 13.6 \\
2442871.0 & KISO & V & $15.1 \pm 0.1$ & 13.5 \\
2443073.3 & KISO & B & $16.7 \pm 0.3$ & 13.7 \\
2443085.2 & KISO & V & $16.1 \pm 0.2$ & 14.5 \\
2443085.3 & KISO & V & $15.9 \pm 0.3$ & 14.3 \\
2443085.3 & KISO & R & $15.2 \pm 0.1$ & 14.4 \\
2443113.2 & KISO & R & $14.29 \pm 0.08$ & 13.49 \\
2443113.3 & KISO & R & $14.25 \pm 0.08$ & 13.45 \\
2443192.9 & KISO & I & $14.48 \pm 0.09$ & 14.48 \\
2443210.0 & KISO & I & $13.62 \pm 0.05$ & 13.62 \\
2443211.0 & KISO & I & $13.63 \pm 0.06$ & 13.63 \\
2443213.0 & KISO & B & $16.5 \pm 0.2$ & 13.5 \\
2443215.0 & KISO & B & $16.6 \pm 0.3$ & 13.7 \\
2443217.0 & KISO & B & $16.7 \pm 0.3$ & 13.7 \\
2443217.0 & KISO & R & $14.3 \pm 0.1$ & 13.5 \\
2443241.0 & KISO & B & $17.5 \pm 0.4$ & 14.5 \\
2443244.0 & KISO & B & $17.2 \pm 0.4$ & 14.2 \\
2443244.0 & KISO & V & $15.8 \pm 0.2$ & 14.1 \\
2443769.3 & KISO & I & $14.42 \pm 0.08$ & 14.42 \\
2443821.0 & KPNO & R & $15.25 \pm 0.09$ & 14.45 \\
2443821.1 & KISO & V & $16.0 \pm 0.1$ & 14.4 \\
2443833.2 & KISO & I & $13.66 \pm 0.06$ & 13.66 \\
2443848.0 & KPNO & I & $13.87 \pm 0.09$ & 13.87 \\
2443949.0 & KPNO & B & $17.4 \pm 0.1$ & 14.4 \\
2444644.0 & KPNO & V & $16.0 \pm 0.2$ & 14.4 \\
2444644.0 & KPNO & R & $15.19 \pm 0.08$ & 14.39 \\
2444644.0 & KPNO & I & $14.43 \pm 0.07$ & 14.43 \\
2444644.0 & KPNO & B & $17.4 \pm 0.1$ & 14.4 \\
2444942.6 & KISO & V & $15.9 \pm 0.2$ & 14.2 \\
2445996.7 & KISO & V & $16.2 \pm 0.2$ & 14.6 \\
2446003.6 & KISO & B & $17.4 \pm 0.6$ & 14.5 \\
2446494.4 & KISO & B & $16.7 \pm 0.2$ & 13.7 \\
2446496.4 & KISO & V & $15.1 \pm 0.1$ & 13.5 \\
2446776.2 & KISO & R & $15.6 \pm 0.2$ & 14.8 \\
2447592.0 & TAU & B & $17.4 \pm 0.2$ & 14.5 \\
2447592.0 & TAU & B & $17.3 \pm 0.2$ & 14.4 \\
2447837.9 & POSS & R & $15.4 \pm 0.2$ & 14.6 \\
2447947.0 & ROZ & B & $17.01 \pm 0.08$ & 14.07 \\
2447947.0 & ROZ & B & $16.89 \pm 0.07$ & 13.95 \\
2447950.0 & ROZ & B & $16.59 \pm 0.08$ & 13.65 \\
2448092.0 & ROZ & B & $17.0 \pm 0.1$ & 14.0 \\
2450015.0 & TAU & B & $17.6 \pm 0.2$ & 14.7 \\
2450365.9 & POSS & I & $ > 15.5 $ & $ > 15.5$ \\
2450432.0 & TAU & U & $18.3 \pm 0.3$ & 14.3 \\
2450481.0 & TAU & U & $18.1 \pm 0.3$ & 14.1 \\
2450482.0 & TAU & B & $17.4 \pm 0.2$ & 14.5 \\
\enddata
\tablenotetext{a}{Effective $I$-band magnitude calculated assuming $U-I=4.02$, $B-I=2.94$, $V-I=1.62$, and $R-I=0.80$}
\end{deluxetable}

\begin{deluxetable}{ccccc}
\tablecaption{Color Measurements of KH15D \label{kh_colors}}
\tablewidth{0pt}
\tablehead{
\colhead{J.D.} &
\colhead{Index} &
\colhead{$C$} &
\colhead{$\Delta C$\tablenotemark{a}} &
\colhead{$\phi$\tablenotemark{b}} 
}
\startdata
  2434801.5 & $B-V$ &  $1.3 \pm 0.2$  & 0.0 & 0.03 \\
  2442747.0 & $R-I$ &  $0.6 \pm 0.1$  & $-0.1$ & 0.26 \\
  2442747.0 & $R-I$ &  $0.9 \pm 0.2$  & 0.1 & 0.26 \\
  2442811.0 & $B-V$ &  $1.5 \pm 0.2$  & 0.2 & 0.58 \\
  2442870.0 & $B-V$ &  $1.5 \pm 0.1$  & 0.2 & 0.80 \\
  2442870.0 & $B-V$ &  $1.4 \pm 0.1$  & 0.1 & 0.80 \\
  2442871.0 & $B-V$ &  $1.6 \pm 0.1$  & 0.2 & 0.83 \\
  2443085.2 & $V-R$ &  $0.9 \pm 0.2$  & 0.1 & 0.25 \\
  2443085.2 & $V-R$ &  $0.7 \pm 0.2$  & $-0.1$ & 0.25 \\
  2443212.0 & $B-I$ &  $2.8 \pm 0.1$  & $-0.1$ & 0.87 \\
  2443217.0 & $B-R$ &  $2.4 \pm 0.2$  & 0.2 & 0.98 \\
  2443244.0 & $B-V$ &  $1.4 \pm 0.2$  & 0.1 & 0.53 \\
  2443821.0 & $V-R$ &  $0.9 \pm 0.1$  & 0.1 & 0.46 \\
  2444644.0 & $B-V$ &  $1.4 \pm 0.2$  & 0.1 & 0.47 \\
  2444644.0 & $V-R$ &  $0.8 \pm 0.2$  & 0.0 & 0.47 \\
  2444644.0 & $R-I$ &  $0.8 \pm 0.1$  & 0.0 & 0.47 \\
  2446495.5 & $B-V$ &  $1.5 \pm 0.1$  & 0.2 & 0.74 \\
  2450481.5 & $U-B$ &  $1.3 \pm 0.4$  & 0.2 & 0.13 \\
\enddata

\tablenotetext{a}{$\Delta C \equiv C_{\rm measured} - C_0$ where $C_0$
is the color index reported by \citet{flaccomio99} (or, in the case of
the single $U-B$ entry, \citet{park00}).}

\tablenotetext{b}{Binary orbital phase, as defined in Eqn.~\ref{eph}.}

\end{deluxetable}

\end{document}